\documentclass{rspublic}
\usepackage{epsf,graphicx,color}
\begin{document}

\title[Effective potentials for polymers and colloids]{Effective
potentials for polymers and colloids: Beyond the van der Waals picture
of fluids?  }  
\author{A.A. Louis} \affiliation{ Department of
Chemistry, Lensfield Rd, Cambridge CB2 1EW, UK }
\label{firstpage}
  \date{\today}
 \maketitle
\begin{abstract}{effective potentials, depletion, colloidal
 suspensions, colloid-polymer mixtures, simple fluids, globular
 proteins}This contribution briefly reviews some recent work
 demonstrating the partial breakdown of the colloidal fluid
 $\leftrightarrow$ atomic fluid analogy.  The success of liquid state
 theory for atomic fluids stems in part from the van der Waals
 picture, where steric interactions dominate the structure, and
 attractive interactions can be added as a perturbation. For complex
 fluids described by effective potentials, this picture may break
 down. In the first example discussed, depletion potentials in
 non-additive hard-sphere mixtures are shown to be surprisingly
 complex, leading to fluid structure and fluid-solid transitions
 dominated by properties of the attractive potentials instead of by
 the hard-cores.  Many colloidal suspensions, and possibly globular
 proteins, fall into this {\em energetic fluid} category.  In the
 second example, the coarse-graining of polymers leads to soft-core
 effective potentials and associated {\em mean field fluid} behaviour
 distinguished by a breakdown of the virial expansion, an equation of
 state that is nevertheless nearly linear in density, and correlation
 functions well described by the random phase approximation.
\end{abstract}

\section{Introduction}

Integrating out a subset of the degrees of freedom
(i.e. coarse-graining) is the first step in many analyses of soft
matter systems.  For colloidal and polymeric suspensions, this
procedure often leads to {\it effective potentials}, and much progress
has emerged from exploiting the analogy between these potentials and
the potentials of atomic and molecular fluids.  In fact, the basic
philosophy behind effective potentials is that the initial effort in
deriving them is recouped when they are input into the well-oiled
machinery of liquid state theory.  However, coarse-graining on the
wide range of length scales available in soft-matter systems leads to
a much richer class of potentials than those found for their atomic
and molecular counterparts (Likos 2001). Since liquid state theory was
originally derived and optimized for the latter, this immediately
implies the possible breakdown of the soft-matter $\leftrightarrow$
atomic fluids analogy.

This paper will focus on a careful derivation of the effective pair
potentials for colloid-colloid, polymer-colloid and pure polymer
suspensions, with a special emphasis on examples where intuition
gleaned from the atomic fluid analogy begins to fail.

In the first example discussed, depletion potentials are derived for
highly asymmetric non-additive binary hard sphere (HS) mixtures. It is
argued that these potentials can describe a much wider class of
asymmetric binary mixtures and lead, in the low density regime of the
larger species, to {\em energetic fluids}, where the structure and
crystallization behaviour is dominated by the effective potentials
rather than  the hard cores. For this reason, a careful derivation of
the effective potentials is particularly important.

The second example concerns modelling polymers as ``soft colloids''.
These are represented by potentials {\em without hard cores}, leading
to the concept of {\em mean-field fluids}, with behaviour quite
different from the HS paradigm that underpins the theory of simple
liquids.

\section{The van der Waals picture of fluids}

But first, let us take a careful look at the theory of simple liquids
(Hansen \& McDonald 1986): Why has it been so successful?  Perhaps the
primary factor is the surprisingly widespread applicability of the
HS fluid as a model of steric effects in atomic and molecular
fluids.  Attractive forces can then be added as a perturbation to a HS
reference system.  This approach is sometimes called the {\it van der
Waals picture} of fluids\footnote{The term van der Waals plus the term
fluid appear in various combinations in the literature. Its use here
follows the definition found in the Chandler, Weeks \& Andersen (1983)
review.}, and a clear summary statement appeared in an influential
review (Chandler {\em et al.\ }1983):

\begin{quotation}
According to the van der Waals picture, the average relative
arrangements and motions of molecules in a liquid (that is the
intermolecular structure and correlations) are determined primarily by
the local packing and steric effects produced by the short-ranged
repulsive intermolecular forces.  Attractive forces, dipole-dipole
interactions, and other slowly varying interactions, all play a minor
role in the structure, and in the simplest approximation their effect
can be treated in terms of a mean-field -- a spatially uniform
background potential -- which exerts no intermolecular force and hence
has no effect on the structure or dynamics, but merely provides the
cohesive energy that makes the system stable at a particular density
or pressure.
\end{quotation}

Historically, the 1957 discovery of a first order freezing transition
in a pure hard-sphere system (Alder \& Wainwright 1957; Wood \&
Jacobson 1957) really got the metaphorical ball rolling, but other
early highlights include:

\begin{itemize}
\item{\bf Freezing}: Longuet-Higgins \& Widom (1964) and Widom (1967)
showed that attractive interactions only mildly perturb the HS
freezing transition near the triple point.
\item {\bf Structure}: Ashcroft and Lekner (1966), and Verlet (1967),
modelled the structure factor $S(k)$ of liquid metals and many other
fluids by the HS $S(k)$ at an appropriate effective HS
diameter. Hansen and Verlet (1969) then connected structure to
freezing by deriving a criterion --- the first maximum peak of $S(k)$
has a value of about $2.8$ near the freezing transition --- that holds
not only for HS fluids, but also for a much wider set of atomic fluids.
\end{itemize}

These ideas were put on a firmer footing by the Barker-Henderson
perturbation theory (Barker \& Henderson 1967), and the more
systematic Weeks--Chandler--Andersen theory of liquids (Chandler \&
Weeks 1970; Weeks {\em et al.\ }1971).  Both theories provide
 ways of performing quantitative calculations for fluids
based on a HS reference system plus a perturbative term arising from
the attractions. That these concepts can also be extended to molecular
fluids with non-spherical hard-core reference systems is illustrated
by a second, more whimsical, citation from the Chandler, Weeks, \&
Andersen (1983) review:
\begin{quotation}
\noindent Similarly the arrangements of molecules in liquid benzene are similar
to the average arrangements of neighboring Cheerios in a bowl of
breakfast cereal, and a solution of argon in benzene should be similar
to the structure achieved when blueberries are mixed with Cheerios.
\end{quotation}

The upshot of all this is that near the triple point, the region for
which liquid state theory was optimised, the structure and
thermodynamics of atomic and molecular fluids is dominated by the
underlying hard-core system; the effect of attractive potentials is
{\em quantitative}, but not {\em qualitative} (at least for
freezing and structure).  This is no longer true for complex fluids.
Instead, as we shall see in the following sections of this paper,
effective potentials may induce {\em qualitative changes} in the
structure and thermodynamics of soft matter systems.  The van der
Waals picture breaks down.

\section{Example A:  Depletion potentials lead to ``Energetic''
Fluids}

\subsection{Depletion potentials for non-additive binary hard-sphere
mixtures}

\begin{figure}
\begin{center}
\includegraphics[width=100mm]{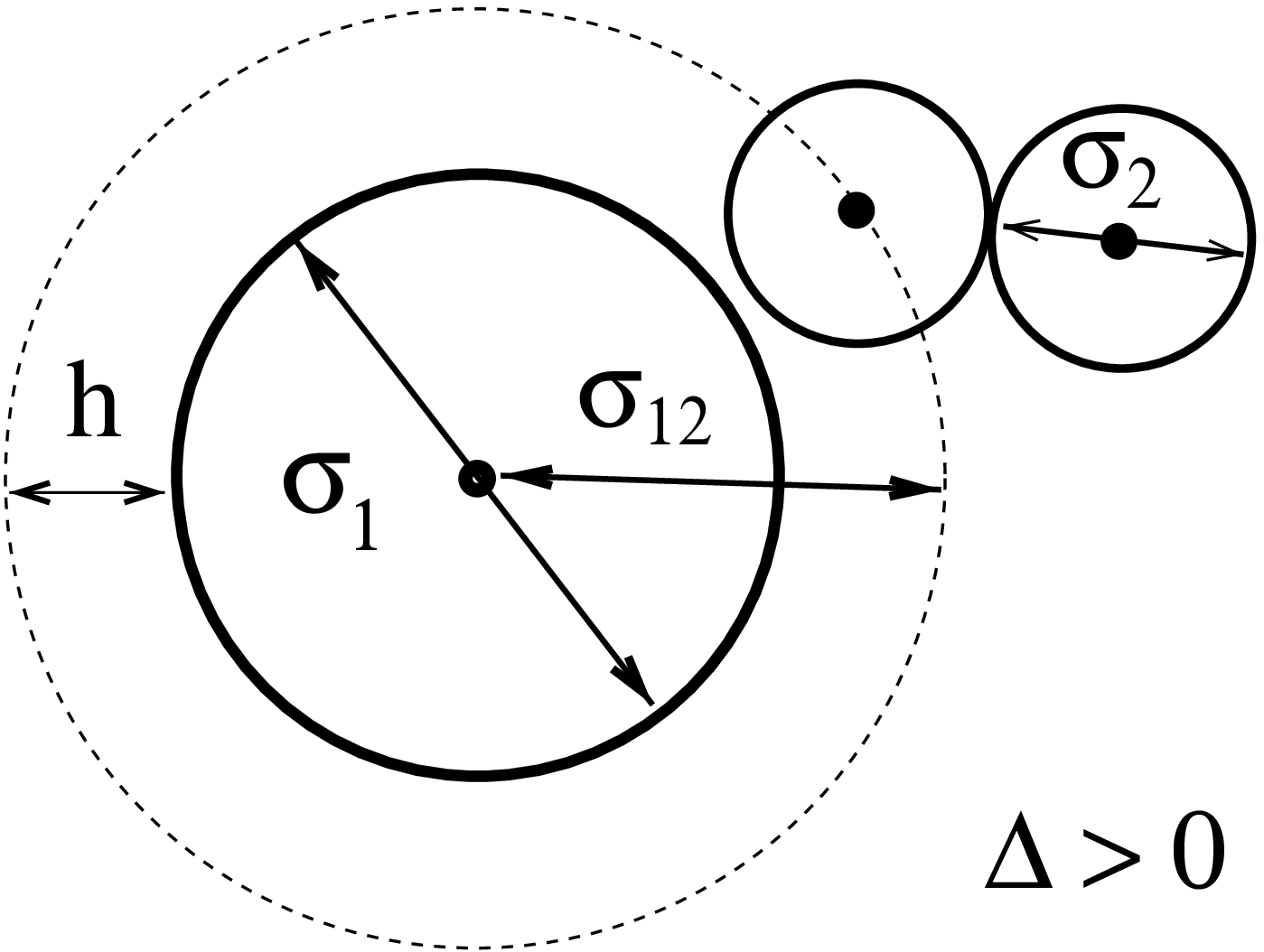}
\begin{minipage}{12cm}
\caption{\label{Fig2.1}The centres of the small spheres can only
approach to within a distance $h=\sigma_{12} - \sigma_1/2 =
\frac{1}{2}\sigma_1(\Delta + y + \Delta y)$ from the surface of the
large spheres ($y=\sigma_2/\sigma_1$).  For an additive system
$\sigma_2 = h$, but here $\sigma_2 <h$ so that the system exhibits
positive non-additivity.  }
\end{minipage}
\end{center}
\end{figure}

Depletion interactions, the effective entropic potentials induced
between the remaining particles when one (repulsive and typically
smaller) component is integrated out, were first described by Asakura
\& Oosawa (1954, 1958).  They remained largely unexplored until around
25 years ago when the experiments of Vincent and co-workers (Li-In-On
{\em et al.\ }1975) and the theoretical work of Vrij (1976) rekindled
interest.  Since then, there has been a steadily increasing body of
work exploiting the ``tunability'' of depletion potentials: the
well-depth is typically proportional to the osmotic pressure of the
small particles while the range is related to their diameters, leading
to rich and interesting experimental phenomena (see e.g. 
Crocker {\em et al.\ }1999;  Poon {\em et
al.\ } 1999; Rudhart {\em et al.\ }1998; Verma {\em et al.\ }1998)
To study this further, consider a binary HS mixture of large (species
1) and small (species 2) spheres.  The complete description of a
binary hard-sphere model demands not only the specification of two
hard-sphere diameters, $\sigma_{1}$ and $\sigma_{2}$, but also the
specification of the cross-diameter $\sigma_{12}$ denoting the
distance of closest approach between two dissimilar spheres. This is
traditionally written as:
\begin{equation}\label{eq1}
\sigma_{12} = \frac{1}{2} \left( \sigma_{1} + \sigma_{2} \right)\,
\left( 1 + \Delta \right)
\end{equation}
When $\Delta =0$, the cross-diameter is simply the sum of the two
 radii, exactly what one would expect on purely geometric grounds; the
 model is termed {\em additive}, and follows the traditional Lorentz
 mixing rule (Hansen \& McDonald 1986).  In nature, however, systems
 will rarely be exactly Lorentz additive, and in fact positive
 ($\Delta > 0$) or negative ($\Delta < 0$) {\em non-additivity} will
 be the rule rather than the exception.  This is further illustrated
 in figure~\ref{Fig2.1}.  Each large particle excludes a volume $v =
 \pi \sigma_{12}^3/6$ from the small particles.  When two large
 particles approach to a distance less than $2\sigma_{12}$, some of
 this volume is doubly excluded, and the small particles can gain
 free-volume, and therefore entropy. Integrating out the small
 particles translates this entropy gain into an interaction between
 the large particles, the so-called depletion potential.  If $2h >
 \sigma_2$, i.e. the effective exclusion diameter of small particles
 near a large particle is larger than their mutual exclusion
 diameters, then $\Delta > 0$ and the system displays positive
 non-additivity.  Conversely, if $2 h < \sigma_2$ then $\Delta < 0$
 and the system displays negative non-additivity.  Experimental
 systems which display negative non-additivity include sterically or
 electrostatically stabilized binary colloid mixtures (Louis {\em et
 al.\ }2000$a$). Polymer-colloid mixtures typically show positive
 non-additivity, and in the extreme limit, $\sigma_{2} \rightarrow 0$
 with $h$ finite, the depletion potential ($w^{(2)}(R_{ij})$ of the
 appendix) reduces to the Asakura-Ooswawa form (Asakura \& Oosawa
 1958, Vrij 1976):
\begin{equation}\label{eq2.1}
\beta V_{AO}(r) = -\rho_2\frac{4 \pi}{3} (\sigma_{12})^3 \left\{ 1 -
\frac{3}{4} \frac{r}{\sigma_{12}} +
\frac{1}{16}\left(\frac{r}{\sigma_{12}}\right)^3 \right\}
\end{equation}
in the range $\sigma_{1} < r \leq 2 \sigma_{12}$; here $\rho_2 =
N_2/V$. The value at contact is given by:
\begin{equation}\label{eq2.2}
\beta V_{AO}(r=\sigma_1) = -\rho_2 \frac{\pi}{4} \left( \sigma_{1}(2
h)^2 + \frac{2}{3} (2 h)^3 \right),
\end{equation}
which does not depend explicitly on the small particle diameter
 $\sigma_2$ or the packing fraction $\eta_2$.

For finite $\sigma_2$ the effective pair potentials reduce to
(\ref{eq2.1}) and (\ref{eq2.2}) only in the $\rho_2 \rightarrow 0$
limit.  Deriving quantitatively accurate depletion potentials at
finite $\sigma_{2}$ and $\rho_2$ for non-additive systems has become
possible due to some important new developments (see e.g. Roth {\em et
al.\ }2000$a$).  Briefly, the method works like this: for a given small
particle fugacity $z_2$, the pair-contribution to the exact effective
potential, $w^{(2)}(r;z_2)$, is given by the difference in grand
potential between a system with two large-spheres at a distance $r$,
and the same system with the two large spheres at $r=\infty$.  Within
a density functional theory (DFT) approach, this can be rewritten as:
\begin{equation}\label{eq2.3}
\beta w^{(2)}(r;z_2) = \lim_{\rho_1 \rightarrow 0} \left(
c_{1}^{(1)}(\infty) - c_{1}^{(1)}(r) \right),
\end{equation}
where $c_{1}^{(1)}(r) = -\beta \delta F_{ex}[\rho_1,\rho_2]/\delta
\rho_1$, $F_{ex}[\rho_1,\rho_2]$ being  the excess intrinsic
free-energy functional of the binary mixture.  By employing the
Rosenfeld fundamental measure theory DFT (Rosenfeld 1989),
quantitatively accurate depletion potentials can be derived for both 
additive (Roth {\em et al.\ }2000$a$) and non-additive mixtures (Roth
\& Evans 2001$b$).  

The Roth DFT method extracts the effective pair-potential from the
one-body correlations as shown in equation~(\ref{eq2.3}). On the other
hand, direct functional differentiation of the same Rosenfeld DFT
leads to two-body correlations that are equivalent to the
Percus-Yevick (PY) approximation in the homogeneous limit (Rosenfeld
1989). But, as shown in the appendix, PY results in a poor
representation of the effective pair potentials.  Curiously, the same
DFT approach provides quantitatively accurate effective potentials by
one route, and rather poor potentials by another route. For this
reason, some care must be taken when choosing a particular route to
thermodynamics or effective potentials from a given (approximate) DFT.

\begin{figure}
\begin{center}
\includegraphics[width=120mm]{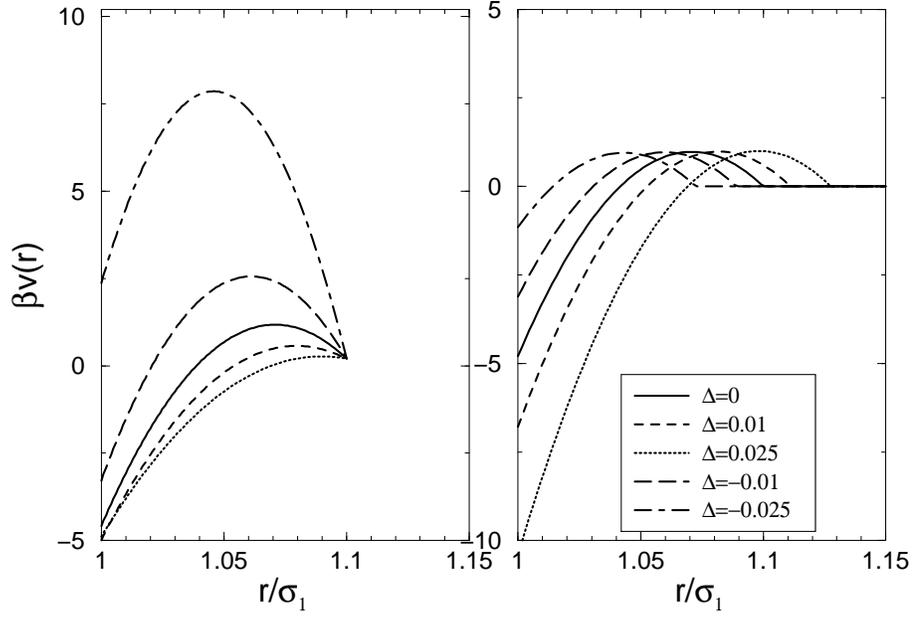}
\begin{minipage}{12cm}
\caption{\label{Fig2.2} The effect of non-additivity on depletion pair
potentials at a size-ratio $y = \sigma_2/\sigma_1 = 0.1$.  In plot (a)
the effective packing fraction $\eta_2^{eff} = 4/3\pi \rho_2 h^3
=0.258045$ is kept constant, while in plot (b) the small-particle
packing fraction $\eta_2 = \pi/6 \rho_2 \sigma_2^3 =0.258045$ is kept
constant. For the AO potential at $y=0.1$, this packing corresponds to
a second virial coefficient $B_2/B_2^{HS}=-1.5$, which is near the
(metastable) fluid-fluid critical point (Vliegenthart \& Lekkerkerker
2000).  The potentials are from Louis {\em et al.\ }(2000a), and
ignore oscillations at a range $r > 2 h$.  Note how the two ways of
varying the non-additivity affect the depletion potentials
differently. }
\end{minipage}
\end{center}
\end{figure}

The effect of non-additivity on the depletion potentials is shown in
figure~\ref{Fig2.2}(a), where the parameters are chosen such that
$\rho_2 h^3$ is kept constant, but $\sigma_{2}$ is varied.  Note in
particular that for positive non-additivity ($\sigma_2 < 2 h$) the
contact value changes very little.  In contrast, for negative
non-additivity ($\sigma_2 < 2 h$), the contact value increases
markedly and can even become positive.  The non-additivity can also be
varied by keeping the small-particle diameter $\sigma_{2}$ and the
packing fraction $\eta_2 = \pi \rho_2 \sigma_2^3/6$ fixed, as shown in
figure~\ref{Fig2.2}(b).  For positive and negative non-additivity, the
oscillations due to the solvation shells remain more or less the same,
but the well depth changes markedly.  These trends can be understood
even at the very simple level of equation~(\ref{eq2.2}), valid only as
$\rho \rightarrow 0$, since changing $\Delta$ while keeping
$\sigma_{2}$ fixed corresponds to changing $h$ and therefore the
contact value.

Non-additivity has an even more dramatic effect on the virial
coefficients.  The additive case seems to be marginal since a very
small negative or positive non-additivity markedly changes the
behaviour of $B_2$ away from the additive value (Louis \& Roth 2000$e$).
A connection to phase behaviour can be made through the recent
observation of Vliegenthart and Lekkerkerker (2000), who showed that
$B_2/B_2^{HS} \approx -1.5$ near the critical point of a wide variety of
fluid systems.  The large effect of non-additivity on the virial
coefficients found by Louis and Roth (2000$e$) can therefore rationalize
the large effect of non-additivity found in previous direct studies of
the fluid-fluid spinodal line (Biben \& Hansen 1997, Dijkstra 1998,
Louis {\em et al.\ }2000$a$).  Positive non-additivity strongly favours
phase-separation, while even a very small negative non-additivity has
the opposite effect.

More realistic two-component systems include attractive or repulsive
potentials $v_{ij}(r)$ in addition to the hard-core steric repulsion.
These can be mapped onto the non-additivity in the following way: An
attractive cross-term $v_{12}(r)$ or a repulsive small-small
interaction $v_{22}(r)$ corresponds to $\Delta > 0$, while a repulsive
$v_{12}(r)$ or attractive $v_{22}(r)$ lead to $\Delta < 0$.  In this
way the large-small or small-small interactions can be used to
``engineer'' a very wide variety of effective potential shapes and
associated fluid behaviours (Louis \& Roth 2000$e$).

\subsection{Energetic fluids: Structure and thermodynamics}

\subsubsection{Thermodynamics}

\begin{figure}
\begin{center}
\includegraphics[width=120mm]{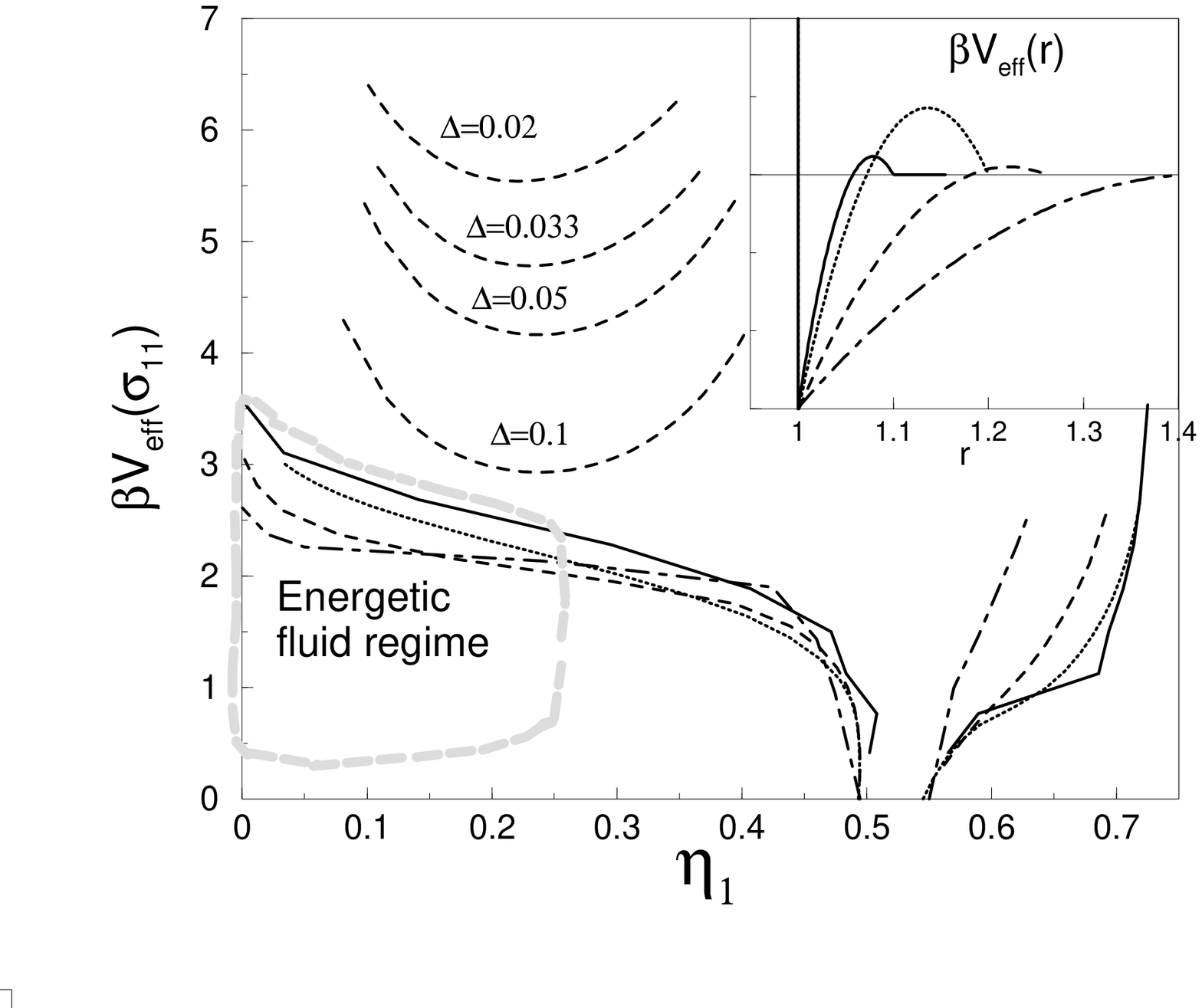}
\begin{minipage}{12cm}
\caption{\label{Fig2.7} Typical phase-diagram for short-range
depletion potential systems.  The meta-stable fluid-fluid lines are
for depletion potentials with $y=0.2$ and different $\Delta$ (Louis
{\em et al.\ }2000a).  The fluid-solid lines come from the potentials
in the inset (each with corresponding line styles), here normalized to
the same minimum at contact.  Note the differences in range and shape
of the potentials.  The area inside the broad dashed lines roughly
denotes the stable energetic fluid regime; above the fluid-solid
curves would be the metastable energetic fluid regime.  The
fluid-solid lines for the 3 shorter range potential systems was
generated with first order perturbation theory, while the $y=0.4$ AO
potential fluid-solid line comes from the simulations of Dijkstra {\em
et al.\ } (1999$b$) because 1st order perturbation theory incorrectly
exhibits a stable fluid-fluid phase-transition.}
\end{minipage}
\end{center}
\end{figure}

Effective depletion potentials can have a much shorter range than the
hard-core diameter of the larger colloidal species.  This leads to
perhaps the best-known breakdown of the simple atomic fluid analogy,
namely the metastability of the fluid-fluid transition w.r.t.\ the
fluid-solid transition for short-range potentials (Gast {\em et al.\
}1983; Hagen \& Frenkel 1994).  As demonstrated in
figure~\ref{Fig2.7}, this leads to crystallization at much smaller
large-sphere packing fractions than found for the archetypical
freezing transition in one-component HS fluids.  The relative
insensitivity of the crystallization line to the form and range of the
potentials shown in figure~\ref{Fig2.7} is particularly striking.  The
determining factor is mainly the depth of the attractive well at
$r=r_{min}$, from which, for the range of potentials probed here, an
approximate crystallization criterion can be derived (Louis {\em et
al.\ }2000$a$): the liquidus line broadens to about $50 \%$ of the
packing fraction at freezing for pure HS's when $\beta
V_{eff}(r_{min}) \approx 2.4 \pm 0.3$\footnote{As the potential
becomes more short-ranged, the trend is towards somewhat higher values
of $\beta V_{eff}(r_{min})$. Nevertheless, this very simple criterion
describes the dominant effect in the range of most interest to
experiment.}

\begin{figure}
\begin{center}
\includegraphics[width=120mm]{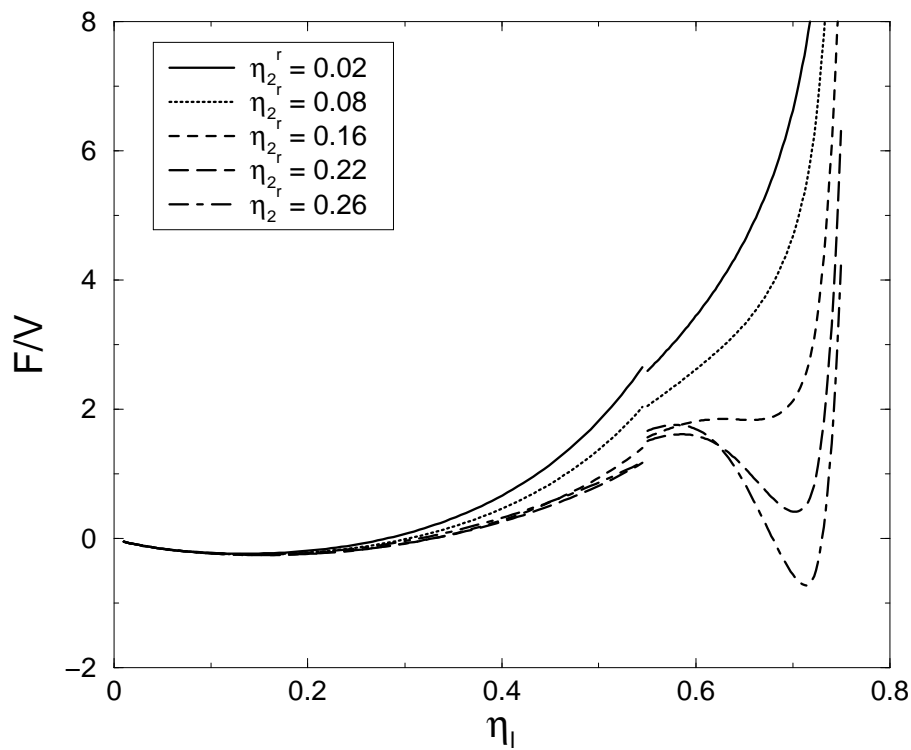}
\begin{minipage}{12cm}
\caption{\label{Fig2.8}Normalized free energies per unit volume for
the additive HS potential used in (Dijkstra {\em et al.\ } 1998), at a
size-ratio $y=0.2$. The packings of the small-spheres are $\eta_2^r =
0.02,0.08,0.16,0.22$ and ,$0.26$, which correspond to well-depths of
$\beta V(r=\sigma_1)=0.183,0.768,1.597,2.224,2.632$, respectively.
Both branches of the free energy curve decrease with increasing
potential well-depth, but the effect is much more pronounced for the
crystal branch.  }
\end{minipage}
\end{center}
\end{figure}

The insensitivity of crystallization to the potential range can be
understood from the trends in the free-energy curve.  With increasing
well-depth, the fluid branch of the free-energy in figure~\ref{Fig2.8}
is only mildly affected, while the solid branch develops a deep
minimum, driven by close contact of the potential wells at a very high
packing fraction\footnote{This well in the crystal free-energy branch
also drives the solid-solid transition}.  Equilibrium between a very
dense solid and a dilute fluid can now easily be achieved. (This can
be seen in figure~\ref{Fig2.8} by using the common tangent
construction.)  
\footnote{While the solid must be treated accurately, small errors in
the fluid-free energy have little effect on the fluid-solid transition
in the energetic fluid regime, in marked contrast to the pure HS case
where both branches of the free energy must be treated accurately. The
free-energies in figure~\protect\ref{Fig2.8} were generated with first
order thermodynamic perturbation theory.  They reproduce the
fluid-solid phase-behaviour of short-range potentials rather well, but
fail miserably for the (metastable) fluid-fluid transition (Dijkstra
{\em et al.\ }1999$b$, Louis 2000$a$). First order perturbation theory
even fails to correctly describe the second virial coefficient, so it
is not surprizing that it breaks down for the fluid-fluid line.}.
Therefore, when the liquidus line broadens, crystallization is driven
primarily by changes in the close-packed solid branch of the free
energy, which probes well depths at the potential minima, but is
largely independent of other features of the potentials.  This is the
origin of the semi-universal crystallization law found by Louis {\em
et al. }(2000$a$).


\subsubsection{Structure}

In the regime where the crystallization curve broadens to lower and
lower packing fractions of the large-particles, the structure can be
well approximated by the very simple form: $g(r) = \exp[-\beta
w^{(2)}(r;z_2)]$ (Louis 2000$b$), where $w^{(2)}(r;z_2)$ is the
two-body contribution of the exact effective potential (EEP) described
in the appendix.  As demonstrated in figure~\ref{Fig2.9}, this works
remarkably well, even for densities as high as $\eta_1 = 0.25$.
Because the fluid-fluid transition is metastable and crystallization
typically sets in only for a well-depth of order $\beta V(r_{min})
\approx 2.4$, the $g(r)$ near contact can easily approach values
greater than $10$ in the fluid phase, which is quite different from
the value for a HS reference system at the same overall density.

 If experiments could directly access $g(r)$, this would lead to a
very simple method to extract the effective potentials.  However,
experiments on pair-correlations typically access $S(k)$, where the
effect of the potentials is much less clear cut, as demonstrated in
figure~\ref{Fig2.10}. In contrast to the van der Waals picture, the
effective potentials alter the $S(k)$ quite markedly from the HS
reference system, but similarly to the van der Waals picture, where
deriving the attractive contribution to the potential from an
inversion of $S(k)$ is very difficult (Reatto 1986), quite a number of
different effective potentials may result in similar behaviour for
$S(k)$.  The potentials in figure~\ref{Fig2.10} were chosen to have
nearly equal virial coefficients $B_2$.  Since $2B_2 = - \hat{f}(0)$,
where $\hat{f}(k)$ is the Fourier transform (FT) of the Mayer function
$f(r) = \exp[\beta v(r)]-1$, (which in turn provides a good
approximation for the total correlation function $h(r)=g(r)-1$ in
this regime) it is not surprizing that $S(k) = 1 + \rho \hat{h}(k)$
should be similar for different potentials as long as $B_2$ is held
constant.\footnote{This simple picture breaks down near the spinodal
line where $S(k=0)$ is enhanced by collective density fluctuations.}.

The fluid-solid transition is mainly determined by the potential
minimum, and is therefore not directly related to $B_2$ (compare, for
example, the virial-coefficients of the potentials depicted in the
inset of figure~\protect\ref{Fig2.7}).  This implies that $S(k)$,
which is mainly determined by $B_2$, will vary significantly for
different potentials along the
fluid-solid line. Therefore the Hansen-Verlet criterion, or any other
similar criterion based on $S(k)$, will not hold in this ``energetic
fluid'' regime.

\subsubsection{The energetic fluid picture vs.\ the van der Waals
picture of fluids}

The term {\em energetic fluid} applies to the fluid phase at low
overall packing fraction of the large-particles (say $\eta_l \leq
0.25$), as shown schematically in figure~\protect\ref{Fig2.7}.  This
regime opens up in short-range potential fluids when the fluid-fluid
phase-line becomes metastable to the fluid-solid line, which then
broadens out to very low packing fractions.  While it is hard to find
atomic or molecular fluids with a potential range short enough to
qualify for this nomenclature, many solutes in solution are governed
by relatively short-range attractive potentials. Through
McMillan-Mayer solution theory (McMillan and Mayer 1945) they can be
viewed as effective atomic fluids. But in contrast to the van der
Waals picture described in section 2, the attractive potentials of
such energetic fluids qualitatively affect both  freezing and structure:
\begin{itemize}
\item {\bf Freezing:} The liquidus line  is determined
primarily by the well-depth of the potential, and less so  by other
 details such as its range or shape.  While in the van der Waals
picture freezing is driven by entropic excluded volume effects, here it
is driven by (energetic) properties of the potentials.
\item{\bf Structure:} The real-space structure of energetic fluids can
be qualitatively described by the simple relation $g(r) = \exp[-\beta
v(r)]$, while the k-space structure $S(k)$ is well described by the
Baxter model at the same $B_2$.  In contrast, the $S(k)$ of van der
Waals fluids typically resembles that of a HS fluid.

is mainly determined by the
value of the second virial coefficient rather than by other details of
the potentials.
\end{itemize}

Examples of energetic fluids include many colloidal suspensions, where
direct interactions can be of short range, or else where other (smaller)
colloids, micelles, or polymers can act as depletants, inducing
potentials such as those depicted in figure~\ref{Fig2.2}.

Globular proteins may also fall into the same class, but their
interactions are no longer  spherically symmetric (Durbin \& Feher
1996; Lomakin {\em et al.\ }1998; Neal {\em et al.\ }1998; Piazza
2000; Sear 1999). The radial distribution function and structure factor
should still display behaviour similar to other energetic fluids, but
now with an implicit orientational average over the anisotropic
potentials.

Proteins typically crystallize at very low packing fractions, and some
show evidence of a phase diagram topology similar to the one depicted
in figure~\ref{Fig2.7} (Frenkel 2000; Piazza 2000; Rosenbaum 1996;
Sear 1999), with a metastable fluid-fluid line. There is also evidence
that potential contacts in the crystal are very important for protein
crystallization (Durbin \& Feher 1996), which is consistent with
energetic fluid behaviour and in sharp contrast to what would be
expected from the van der Waals picture.  The semi-universality of
crystallization found for spherically symmetric potentials depends on
similar behaviour of the solid branch of the free-energy for different
potential shapes.  Unfortunately, this can no longer be expected for
proteins, since (a) the proteins are no longer spherical and (b) the
potentials are anisotropic (or ``patchy''), leading to many possible
crystal structures.  In fact, as emphasized by Sear (1999), conditions
(a) and (b) suggest that some proteins may be very difficult to
crystallize precisely because their potential contacts are not
commensurate with an allowed crystal structure. So, whether globular
proteins can be usefully classified as energetic fluids remains to be
seen, but they can certainly not be classified within the van der
Waals picture of fluids.

\section{Example B: Polymers as soft colloids:  Mean field fluids}


\subsection{Deriving effective polymer-polymer potentials}

The approach outlined here, described in more detail in (Louis {\em et
al.\ }2000$c$, Bolhuis {\em et al.\ }2001), is to coarse-grain the
polymers by integrating out monomeric degrees of freedom, resulting in
a description based on the polymer centres of mass (CM).  In
principle, for a set of $N_1$ polymers with $L$ monomers each, one
could calculate the EEP which depends explicitly on the configuration
$\{R_i\}$ of the centres of mass.  The only differences with the
binary-mixture case described in the appendix are: (1) there is no
explicit additional dependence on the monomer density (or fugacity),
since that is fixed by the number of polymers in the set $\{R_i\}$,
and (2) there is no {\em direct} interaction $V(\{R_i\})$; the
polymer-polymer effective potential arises solely from the
coarse-graining procedure.  The EEP now takes the form:
\begin{equation}\label{eq3.1}
W^{eff}(\{R_i\}) = \sum_{i<j}^{N_1}  w^{(2)}(R_{ij})  +
\sum^{N_1}_{i<j<k}  w^{(3)}(R_{ijk})  + \ldots,
\end{equation}
and again, each term could in principle be calculated separately, but
this rapidly becomes intractable.  Instead, one can follow the second
route described in the appendix and include all higher order terms in
equation~(\ref{eq3.1}) through the density dependence of an effective
pair potential $w(r;\rho_1)$, which exactly reproduces the
CM pair-correlations at a given density $\rho_1 = N_1/V$.  To do so, we
first perform Monte Carlo (MC) simulations of self avoiding walk
$L=500$ chains on a simple cubic lattice and extract the
radial-distribution functions $g(r)$ between the CM.  These are then
inverted through an Ornstein-Zernike procedure using the
hypernetted-chain closure (HNC), which has been shown to be
quasi-exact for the potentials under consideration (Louis {\em et al.\
}2000$b$, 2000$d$, Bolhuis {\em et al.\ }2001).  Figure~\ref{Fig3.1} 
shows both the radial-distribution functions and the effective
potentials inverted from the $g(r)$.  According to the theorems
discussed in the appendix, the effective-potentials should, through
the pair-correlations and equation~(\ref{eqA.13}), reproduce the
contributions to the (osmotic) compressibility induced by the
EEP~(\ref{eq3.1}).  Figure~\ref{Fig3.2} shows that the total
compressibility is very well approximated in this approach, implying
that, in contrast to the binary fluid case, the volume term
contribution (defined in the appendix) is small. The difference in
volume terms is most likely due to the fact that the number of
monomers is fixed by the number of polymers, instead of being free to
vary as is the case for the small particles in the binary mixture. In
fact, if the polymers were rigid, their CM correlations would exactly
determine the compressibility (see e.g.\ Hansen \& McDonald 1986 ch 12).

\subsection{Mean Field Fluids: structure and thermodynamics}

The effective potentials shown in figure~\ref{Fig3.1} are radically
different from the usual hard-core plus attraction form found for
atomic systems, immediately suggesting different fluid behaviour.
Already the radial distribution functions in figure~\ref{Fig3.1}
appear to be quite different from their atomic-fluid counterparts.
The structure of fluids described by these potentials is well
approximated by the simple random phase approximation over a
surprizingly large density range (Louis {\em et al.\ }2000$d$, Likos
{\em et al.\ }2000).  Marked differences with atomic fluid behaviour
also arise at the level of thermodynamics, where the equation of state
(EOS) is very well described by the following mean-field form:
\begin{equation}\label{eq3.2}
Z = \frac{\beta P}{\rho} \approx Z_{MF} = 1 + \frac{1}{2}
\rho \int d{\bf r} \beta v(r) = 1 + \frac{1}{2} \beta \hat{v}(k=0) \rho
\end{equation}
As shown in figure~\ref{Fig3.3}, this also hold for integrable potentials which
diverge at the origin.  Equation~(\ref{eq3.2}) demonstrates that what
matters is the value of the FT of the potential at the origin,
$\beta \hat{v}(k=0)$, which is finite for all 3 potentials, and, in
figure~\ref{Fig3.3}, chosen to have the same value to ensure that the
three EOS are very close.  The quasi-linear behaviour of the EOS
resembles that of an atomic fluid in the second virial approximation,
but, in fact, the three potentials have virial coefficients differing by
about $10\%$.  More seriously, the virial expansion has a surprisingly
small radius of convergence so that adding higher order virial
coefficients results in a much poorer approximation (Louis {\em et
al.\ }2000$d$), as demonstrated by the third order virial expansions in
figure~\ref{Fig3.3}.

Instead of the CM, end-points or mid-points could also be used to
construct an effective particle picture of interacting polymer
solutions.  For example, the mid-point representation would be very
similar to the two-arm limit of a star-polymer, for which a number of
results have been recently derived (Likos {\em et al.\ }1998; Likos
2001; Watzlawek {\em et al.\ }1999).  In figure~\ref{Fig3.4} the $f=2$
limit of the star-polymer potential is compared to a more recent
expression for the mid-point--mid-point interaction (Dzubiella {\em et
al.\ }2000).  At first sight the two appear very similar, but the
slope of their respective EOS differ by a factor of four!  The reason
for this can be seen in figure~\ref{Fig3.5}: Since $\beta
\hat{v}(k=0)$ is proportional to the integral over $r^2 \beta v(r)$,
it is the small differences (much less than $k_B T$) in the tails of
the two $\beta v(r)$ that determine the large differences in the EOS.
Interestingly the two radial-distribution functions are again quite
similar (although the associated structure factors are not).

\subsection{Mean field fluids vs.\ the van der Waals picture of fluids}

A necessary, but not sufficient condition for mean field fluid (MFF)
behaviour is that $\beta \hat{v}(k=0)$ is finite.  If $\beta
\hat{v}(k=0)$ is relatively large, then MFF behaviour, with its
quasi-linear EOS and correlation functions well-described by the
random phase approximation, only sets in for large enough densities.
At lower densities the structure can still resemble that of a HS
fluid, but the topology of the phase-diagram is quite different (Lang
{\em et al.\ }2000, Louis {\em et al.\ }2000$d$).  Depending on
whether the FT of the potential oscillates or not, such fluids may
show either re-entrant melting or else a clustering transition for
large enough values of $\beta \hat{v}(k=0)$ (Likos {\em et al.\
}2000).

Systems with values of $\beta \hat{v}(k=0)$ significantly below the
value that can induce freezing or clustering transitions are MFF at
all densities. The associated simple equation of state and
fluid-structure are markedly different from the behaviour of fluids in
the van der Waals picture\footnote{Perhaps I'm beating a straw man
here, but the distinctions are hopefully helpful nonetheless}.

\section{Conclusions}

While the van der Waals picture of fluids lies at the basis of many
successful liquid state theories for atomic and molecular fluids, it
breaks down for the richer class of potentials arising from 
coarse-graining procedures in complex fluids.  Nevertheless, a judicious
choice of techniques drawn from the theory of simple liquids may still
provide insight when combined with a careful derivation of effective
potentials.  Some recent examples of this programme include the
phase-diagram of additive (Dijkstra {\em et al.\ }1999$a$) and
non-additive (Louis {\em et al.\ }2000$a$) asymmetric binary HS
mixtures, the structure and phase-behaviour of star-polymers (Likos
{\em et. al.\ }1998; Watzlawek {\em et. al.\ }1999), and star-polymer
colloid mixtures (Dzubiella {\em et. al.\ }2000), and the structure and
phase-behaviour of a pure-polymer system (Louis {\em et. al.\ }2000$c$;
Bolhuis {\em et. al.\ }2001).  In each case the effective potential
picture was the key to clarifying the underlying physics.

Similarly, in this contribution a derivation of the effective
potentials in binary hard-sphere fluids demonstrated the key role of
non-additivity in determining the shape of the effective depletion
potentials, and the associated phase behaviour.  When such potentials
are short ranged, they open up a region in the phase-diagram where the
fluid behaviour deviates significantly from the van der Waals picture,
so much so in fact, that a new nomenclature, {\em energetic fluids},
has been introduced.  In the energetic fluid regime, the structure can
be described by a very simple form: in real space $g(r)=\exp[-\beta
w^{(2)}(r;\rho_2)]$, while in reciprocal space $S(k)$ is largely
determined by the virial coefficient $B_2$. For a wide variety of
potential shapes, the liquidus line broadens to low large-particle
densities when well-depth $\beta w^{(2)}(r_{min};\rho_2) \approx
2.4$.

When linear or star polymers in solution are represented as ``soft
colloids'' centred around their mid-points or their CM, the resultant
picture leads to effective potentials with a finite value of $\beta
\hat{v}(k=0)$.  Again, such fluids do not follow the van der Waals
picture.  For example, their EOS closely follows a mean-field linear
form $\beta P/\rho \approx 1 + 1/2 \rho \beta \hat{v}(k=0)$, and their
structure is well described by the simple random phase approximation
closure: behaviour best classified under the moniker {\em mean field
fluids}.

\begin{acknowledgements}
Most of this work was done in close collaboration with J.P. Hansen,
P.G. Bolhuis, R. Roth, R. Finken, and E.J. Meijer, and has appeared in
our joint publications listed in the references.  I thank David Rowan
for a critical reading of the manuscript and gratefully acknowledge
financial support from the Isaac Newton Trust, Cambridge.
\end{acknowledgements}

\appendix[]{Two ways of deriving effective  potentials for binary mixtures}

Consider a binary mixture of large (species $1$) and small (species
$2$) spherical particles.  Integrating out the smaller component to
derive a new one-component fluid interacting through effective
depletion interactions is a useful way to treat such systems.  This
appendix will focus on two popular ways to derive these effective
potentials:

\subsubsection{Method 1:  Exact effective potential}

For binary mixtures the exact effective potential (EEP) is most easily
analyzed in the semi-grand ensemble, where the number of large
particles $N_1$ and the fugacity $z_2$ of the small particles is fixed
(Lekkerkerker {\em et al.\ }1992, McMillan \& Mayer 1945)\footnote{
Charged systems are best analyzed in the the canonical ensemble. An
early example would be the effective potentials in liquid metals
(Ashcroft \& Stroud 1978)}.  Given a set of $N_1$ large particles
fixed at positions $\{R_i\}$ in a volume $V$, the small particles are
integrated out by calculating their partition function in the fixed
external field generated by the large particles.  This results in an
effective grand potential for the small particles of the form:
\begin{equation}\label{eqA.1}
\Omega (N_1, z_2, V; \{R_i\}),
\end{equation}
which depends parametrically on the large-particle positions
$\{R_i\}$\footnote{Explicit temperature dependence is omitted in this
section}.  To make further progress we follow the analysis of Dijkstra
{\em et.\ al} (1999$a$, 1999$b$, 2000), and decompose this grand
potential into n-body terms.  Using this mapping, a one-component
system can be derived with an an effective interaction of the form:
\begin{equation}\label{eqA.2}
v_{11}(\{R_i\}) + \Omega (N_1, z_2, V; \{R_i\}) =
V^{(0)}(N_1, z_2, V) + W^{eff}(\{R_i\};z_2),
\end{equation}
where $v_{11}(\{R_i\})$ is the direct interaction between the large
spheres that is already present in the original two component system,
$V^{(0)}(N_1, z_2, V)$ is the so-called volume term (which depends on
the set $\{R_i\}$ only through the total number of large particles
$N_1$, but is independent of their relative positions)\footnote{For
binary fluids the volume term is given by: $V^{(0)}(N_1, z_2, V) = -
p_2(z_2) V + N_1 w^{(1)}(z_2)$ where $p_2$ is the pressure of a pure
component $2$ system at fugacity $z_2$, and $w^{(1)}(z_2)$ is the
grand potential difference due to adding one large component $1$
particle to the same system. One body terms which depend on the
configuration $\{R_i\}$ are zero due to translational invariance.},
and $W^{eff}(\{R_i\};z_2)$ is the EEP which can be further expanded
as:
\begin{equation}\label{eqA.3}
W^{eff}(\{R\};z_2) = \sum_{i<j}^{N_1} \left(v_{11}^{(2)}(R_{ij}) +
w^{(2)}(R_{ij};z_2) \right) + \sum^{N_1}_{i<j<k}
 \left(
v_{11}^{(3)}(R_{ijk}) + w^{(3)}(R_{ijk};z_2) \right)
+ \ldots
\end{equation}
The pair term is simply the sum of the pair contribution to the direct
interaction, $v_{11}^{(2)}(R_{ij})$, and the effective interaction
$w^{(2)}(R_{ij};z_2)$ induced by integrating out the small spheres.
It is precisely this pair term which is measured in the laser tweezer
experiments (Verma {\em et al.\ }1998, Crocker {\em et al.\ }1999).
The triplet term is a similar sum of direct and induced terms with
$R_{ijk}$ the standard 3-body coordinates and so forth for n-body
interactions. (Since the individual terms $w^{(n)}(R_{ij...})$ depend
only on the large-small and small-small interactions, their form is
independent of $v_{11}(\{R_i\})$, which need not be decomposable as a
sum of pair terms.)  The induced pair term can be rigorously defined
as the difference between the grand potential of
equation~(\ref{eqA.1}) for two particles a distance $R_{ij} = |{\bf
R_i}-{\bf R_j}|$ apart, and the grand potential when the two particles
are an infinite distance apart.  From (\ref{eqA.3}) this reduces to
$w^{(2)}(R_{ij};z_2) = \Omega(N_1=2,z_2,V; R_{ij}) - V^{(0)}(N_1=2,
z_2,V)$.  Similarly the induced three-body potential can be defined
as:
\begin{eqnarray}\label{eqA.4}
w^{(3)}(R_{ijk};z_2) & = & \Omega(N_1=3,z_2,V;R_{ijk}) - V^{(0)}(N_1=3,
z_2,V) \nonumber \\
& -& w^{(2)}(R_{ij};z_2)  - w^{(2)}(R_{ik};z_2)  - w^{(2)}(R_{jk};z_2),
\end{eqnarray} i.e. it is that part of the interaction induced by 3 large
spheres which cannot be described by volume and pair interaction terms
alone.  By continuing in similar fashion for higher and higher order
terms the full EEP can be built up such that the
exact free energy of the mixture is given by:
\begin{equation}\label{eqA.5}
F(N_1,z_2,V) = V^{(0)}(N_1,z_2,V) + F^{eff}(N_1,z_2,V),
\end{equation}
where $F^{eff}(N_1,z_2,V)$ is defined as the free energy of a
one-component system interacting through the EEP,
i.e. $\exp[-\beta F^{eff}] = Tr_1 \exp[-\beta W^{eff}]$.  In other
words, by exactly integrating out the small particles for an arbitrary
configuration $\{R_i\}$, the original two-component partition sum has
been rewritten as a weighted sum over large particle
configurations only.  The EEP describes the weighting
of each individual large-particle configuration in the effective
one-component partition sum, while the volume term adds the left over
contributions independent of the large-particle configurations.

Calculating the EEP to all orders is usually impractical. Instead what
is often done is to truncate the series (\ref{eqA.3}) and retain only
the pair potential.  In many cases this is not such a bad
approximation, and for a few systems it is even exact.  For example,
the EEP for the Asakura Oosawa (AO) model (Asakura \& Oosawa 1958,
Vrij 1976)
with size-ratio $q = \sigma_2/\sigma_1 < 2/\sqrt{3}-1 \simeq 0.1547$
is exactly described by the pair term\footnote{Another even simpler
example with only a pair term is the exactly solvable lattice model of
Frenkel and Louis (1992).}.  Similarly, for $ 0.1547 < q < \sqrt{3/2}-1
\simeq 0.2247$ the EEP is exactly described by the pair term and a
triplet term.  But even for this rather simple model, the exact form
of $w^{(3)}(R_{ijk};z_2)$ is very tedious to calculate (Goulding
2000).  Thankfully, simulations have shown that neglecting the higher
order terms for the AO model often works admirably well even for size
ratios as large as $q=0.5$, (Meijer \& Frenkel 1994; Dijkstra {\em et
al.\ }1999$a$, 1999$b$).

\subsubsection{Method 2: Inverting the pair correlations}

The correlations functions in an effective one-component system
interacting through the full EEP (\ref{eqA.3}) are equal to the
correlation functions between the large particles in the original
two-component system, as long as both are at the same state point (see
e.g. Dijkstra {\em et al.\ }2000)\footnote{The volume terms do not
directly contribute to the pair and higher order correlations although
they may contribute indirectly through changing the phase-behaviour of
the system.  For the binary mixtures considered here they don't affect
phase-boundaries (Dijkstra {\em et al.\ }1999$a$), but for other
systems they can (van Roij \& Hansen 1997; van Roij {\em et al.\
}1999; Graf \& L\"{o}wen 1998; Warren 2000), and so must be taken into
account to correctly describe the equivalent state points of the
original two-component and effective one-component systems}.  For
example, $S(k)=S_{11}(k)$, where the first structure factor is that of
the effective one-component system and the second is that of the
original two-component system.  This equivalence also implies that
\begin{equation}\label{eqA.10}
\lim_{\rho_1 \rightarrow 0} g_{11}(r;z_2) = \exp\left[ - \beta
\left(v_{11}^{(2)}(r) + w^{(2)}(r;z_2)\right)\right],
\end{equation}
where $g_{11}(r;z_2)$ is the radial distribution function of the
large-spheres. \footnote{Most information about $g_{12}(k)$ and
$g_{22}(k)$ is lost in the mapping.}.  Similar expressions can be
derived to link higher order correlation functions to higher order
terms in the EEP (\ref{eqA.3}).  Hence, there is a {\em direct link}
between the EEP of the effective one-component system and the $\rho_1
\rightarrow 0$ limit of the fluid correlation functions in the
homogeneous phase of the original two-component model\footnote{Note
that equation \protect\ref{eqA.10} is different from the potential of
mean force which is typically defined as $\beta w^{pmf}(r) =
-\ln[g(r)]$ for any density and is strictly speaking not a pair
potential, but a restatement of the pair-distribution function.  For
example, a system directly interacting through a $\beta w^{pmf}(r)$
derived at finite $\rho$ will not have the same correlations as the
original system from which the potential of mean force was derived. }

The relation (\ref{eqA.10}) may be generalized to finite densities of species
1,  because there exists a one--to--one mapping:
\begin{equation}\label{eqA.11}
g(r;\rho) \leftrightarrow w(r;\rho)
\end{equation}
between a given pair distribution function $g(r)$ at density $\rho$
and a {\em unique} two-body pair potential $ w(r;\rho)$ which
reproduces $g(r)$ {\em irrespective of the underlying many-body
interactions} in the system (Henderson 1974; Chayes {\em et al.\
}1984).  Since at finite densities the radial distribution function
$g(r;\rho_1,z_2)=g_{11}(r;\rho_1,z_2)$ includes contributions not only
from the pair-interactions, but also from higher order contributions
to $W^{eff}(r;z_2,\{R_i\})$, the effective pair-potential includes
these terms in an averaged way.  It can  be written as:
\begin{equation}\label{eqA.12}
w(r;\rho_1,z_2) = v_{11}(r) + \tilde{w}(r;\rho_1,z_2)
\end{equation}
which is connected to equation~(\ref{eqA.10}) through $\lim_{\rho_1
\rightarrow 0} \tilde{w}(r;\rho_1,z_2) = w^{(2)}(r;z_2)$.

The price paid for including the effect of all higher order terms is
to introduce a density dependence in the pair-potential, but the
payoff is that the pair-correlations are exactly reproduced (but not
the triplet or higher order correlations).  Thermodynamics can then be
extracted through the compressibility relation:
\begin{equation}\label{eqA.13}
\left( \frac{\partial \beta  \Pi_1}{\partial \rho_1}\right)_{N,z_2}
=\lim_{k\rightarrow 0}\frac{1}{S(k)}.
\end{equation}
Note, however, that volume terms can also contribute to the total EOS
(Dijkstra {\em et al.\ }2000), so that equation~(\ref{eqA.13})
describes the osmotic compressibility due to species 1 only.  In some
cases this is only a small fraction of the total compressibility of
the full two-component system (Louis {\em et al.\ }1999; Dijkstra {\em
et al.\ }2000).

How does one perform the $g(r;\rho) \leftrightarrow w(r;\rho)$ inversion?
If the full EEP includes only pair terms (as is the case for the AO
model with $q < 0.1547$), then $w(r;\rho_1,z_2) = w^{(2)}(r,z_2)$ at
all densities.  If it only includes pair and triplet terms (as for the
AO model with $ 0.1547 \leq q \leq 0.2247$) then to a good
approximation (Reatto \& Tau 1987; Attard 1992):
\begin{equation}\label{eqA.14}
w(r_{12};\rho_1,z_2) \approx w^{(2)}(r_{12};z_2) -\rho_1 \int d{\bf
r_3}\left[ \re^{-w^{(3)}({\bf r_1},{\bf r_2},{\bf r_3};z_2)} -1
\right] g(r_{13})g(r_{23}).
\end{equation}
More generally, one needs (i) a method to generate the exact
$g_{11}(r)$ for the two component system and (ii) an inversion method
to extract $\beta v(r)$ from $g(r)$.  Inversion methods based on the
Ornstein-Zernike relations exist (Reatto 1986, Zerah and Hansen 1986),
but these are very sensitive to the underlying approximations and the
quality of the original pair-correlations used as input.

As an example of the difficulties involved in (i), consider the popular Percus
 Yevick (PY) approximation which is exactly soluble for binary HS
 mixtures (Lebowitz \& Rowlinson 1965).  On the one hand PY
 approximates the EOS very well, but on the other hand the PY
 approximation to the large-particle correlation function in the
 $\rho_1,\rho_2 \rightarrow 0$ limit reduces to:
\begin{equation}\label{eqA.15}
\lim_{\rho_1,\rho_2 \rightarrow 0}g_{11}^{PY}(r) = 1 - \beta V_{AO}(r)
\end{equation}
instead of the correct exponential form: $\exp[ -\beta V_{AO}(r)]$
(note that $w^{(2)}(r;\rho_2) = V_{AO}(r)$ in this limit).  For finite
$\rho_2$, $w^{(2)}(r;\rho_2)$ begins to deviate from the AO form, but
PY still approximately linearizes the exponential, as illustrated in
figure~\ref{FigA.1}.  In spite of the fact that PY successfully
describes the EOS, it clearly fails quite badly for the effective
potential, especially near contact\footnote{PY is reasonably accurate
for the EOS because it is dominated by the small particle
contributions in the $\rho_1 \rightarrow 0$ limit.}, suggesting that
an inversion of PY at finite densities should also fail.  More
generally, using 2-component integral equations to derive effective
pair potentials or phase-behaviour in the ``colloidal limit'' (small
$y$, large $\eta_2$ and small $\eta_1$) is fraught with difficulty.
For example, Biben {\em et al.\ }(1996) showed how two self-consistent
closures that work extremely well for one-component systems, RY and
BPGG, predict quite different locations of the fluid-fluid spinodal
line in binary HS mixtures. A comparison of the effective pair
potentials $w^{(2)}(r;\rho_2)$ calculated for each closure by these
authors can, in fact, rationalize the difference, since these
effective potentials are what primarily determines the
phase-behaviour.  This suggests that for two-component integral
equations in the colloidal limit, it is much better to compare the
performance for the large particle osmotic pressure rather than the
performance for the full EOS, which is typically dominated by the
small-particle contribution.

\newpage

\begin{figure}
\begin{center}
\includegraphics[width=120mm]{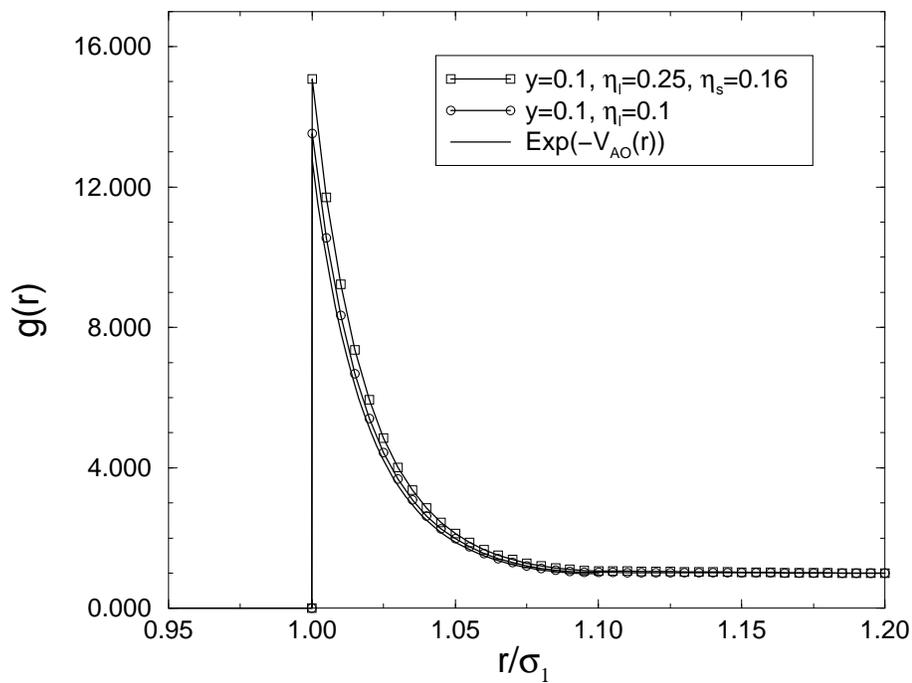}
\begin{minipage}{12cm}
\caption{\label{Fig2.9} The exponential form $g(r) = \exp(-\beta
V_{AO}(r)$ is a semi-quantitative approximation for these two
state-points in the energetic fluid regime.  When $\eta_l = 0.25$, the
system is near the fluid-solid line.  The $g(r)$ were generated with
the Percus Yevick (PY) approximation, which is quantitatively accurate
in this regime (see e.g. Dijkstra {\em et al.\ } 1999$b$)).  }
\end{minipage}
\end{center}
\end{figure}

\begin{figure}
\begin{center}
\includegraphics[width=120mm]{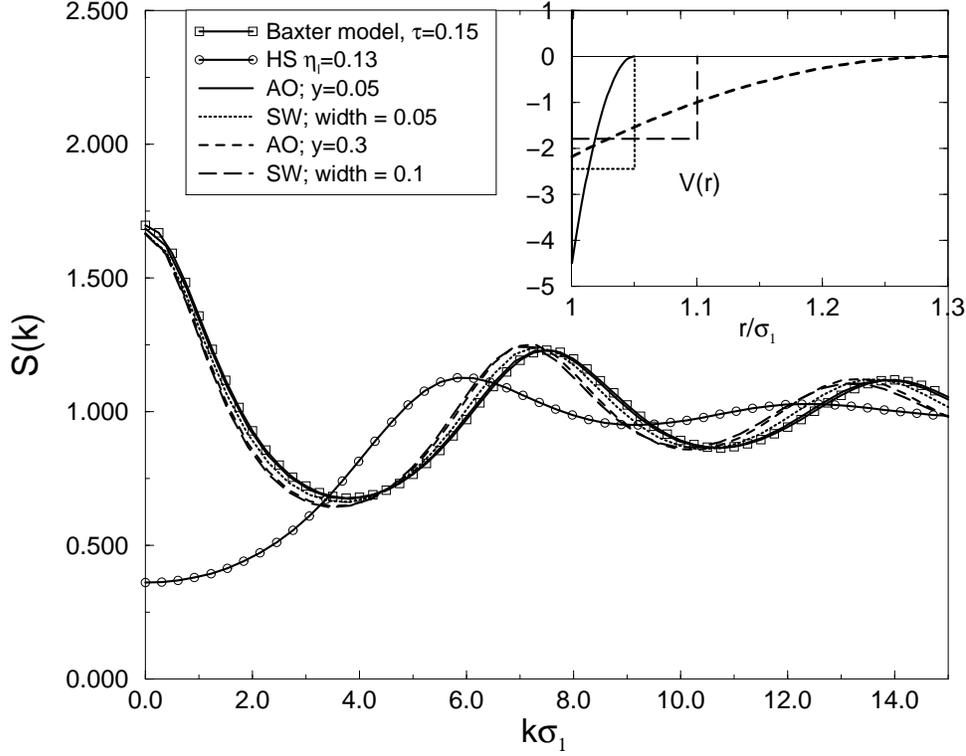}
\begin{minipage}{12cm}
\caption{\label{Fig2.10} The structure-factors $S(k)$ at $\eta_l =
0.13$ for the potentials in the inset.  Note that the shape of $S(k)$
is poorly approximated by the pure hard-sphere form, but well
represented  by the Baxter model $S(k)$.  The two shorter range
potentials have the same $B_2$ as the Baxter model, while the two
longer range potentials have a $B_2$ about $7\%$ more negative, which
fits to almost the same $S(0)$.  The differences in the $S(k)$ due to
the shape and range of the potentials are not much larger than the
resolution of the best experiments on colloids in solution.  As in the
previous figure, the $S(k)$ were determined by the accurate PY
approximation.}
\end{minipage}
\end{center}
\end{figure}

\begin{figure}
\begin{center}
\includegraphics[width=120mm]{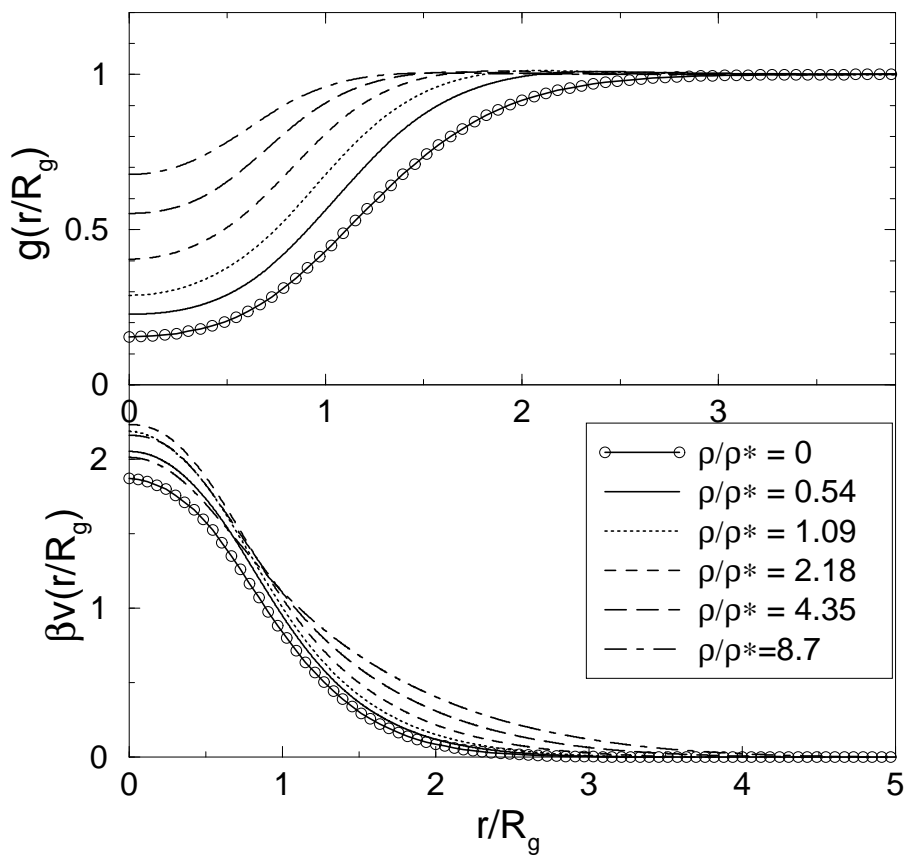}
\begin{minipage}{12cm}
\caption{\label{Fig3.1} MC simulations are used to generate the $g(r)$
for $L=500$ SAW polymer chains on a cubic ($240^3$) lattice for
densities from $\rho \approx 0$ ($N=2$ polymers) to $\rho/\rho^*=8.7$,
($N=6400$) Here $\rho^* = 4/3\pi R_g^3$, and $\rho/\rho^* < 1$ denotes
the dilute and $\rho/\rho^* > 1$ the semi-dilute regimes for polymers
in a good solvent. These radial distribution
functions are inverted using an Ornstein-Zernike procedure to obtain
the effective potentials between the polymer CM.}
\end{minipage}
\end{center}
\end{figure}

\begin{figure}
\begin{center}
\includegraphics[width=120mm]{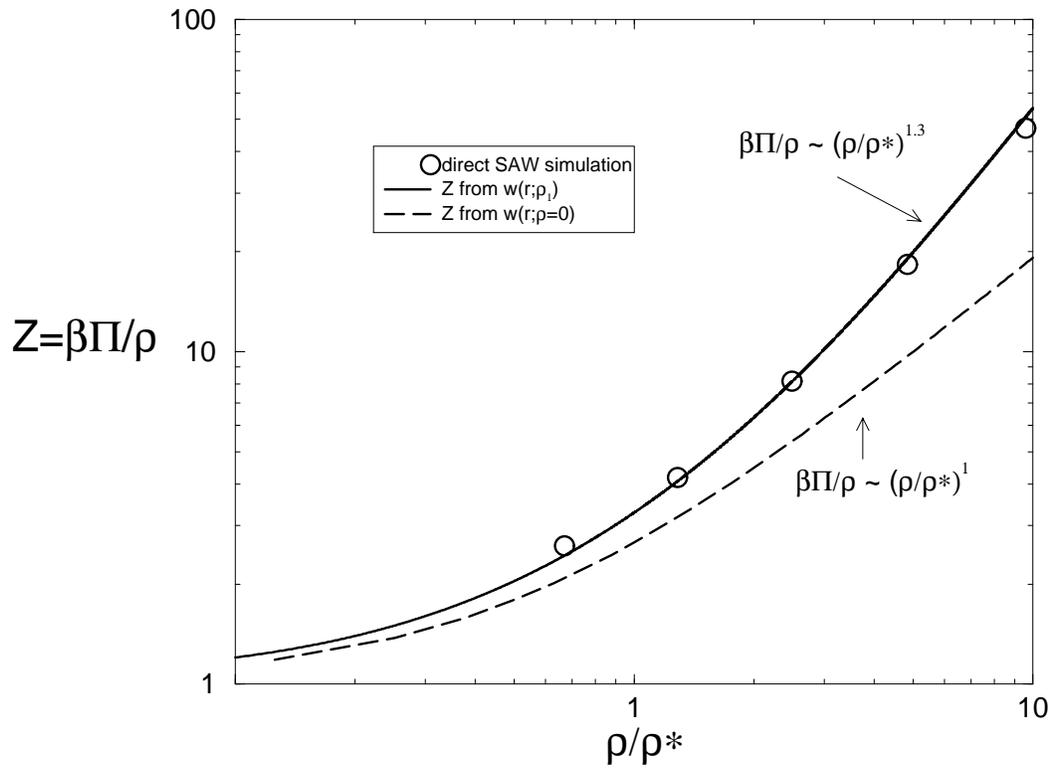}
\begin{minipage}{12cm}
\caption{\label{Fig3.2} The equation of state derived from the
compressibility relation~(\protect\ref{eqA.13}) accurately
approximates the true EOS measured by direct simulations of the
polymers.  This implies that the volume terms are small.  Ignoring the
density dependence of the potentials by using only the low density
form of the pair potential, $w(r;\rho=0)$, equivalent to the
pair-contribution to the EEP, leads to an underestimate of the
pressure}
\end{minipage}
\end{center}
\end{figure}

\begin{figure}
\begin{center}
\includegraphics[width=150mm]{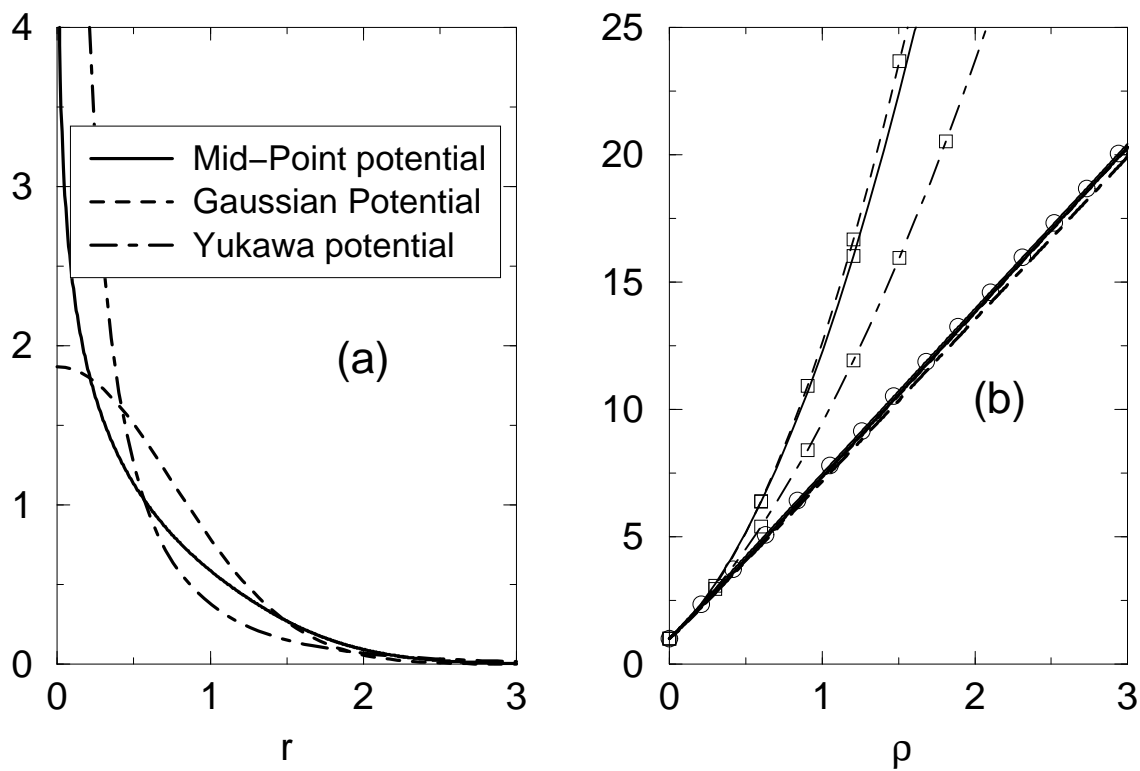}
\begin{minipage}{12cm}
\caption{\label{Fig3.3} (a): Three potentials $\beta V(r)$ which all
result in a mean field fluid.  (b): The EOS $Z=\beta P/\rho$, here
generated by the quantitatively accurate HNC approximation, are all
very close to the $Z_{MF}$ (equation~(\protect\ref{eq3.2})) form
(circles).  However they are badly approximated by a third order
virial series $Z = 1 + B_2 \rho + B_3 \rho^2$, here represented by
squares along each line. In this and the next two plots, line styles in
(a) and (b) denote corresponding systems.  }
\end{minipage}
\end{center}
\end{figure}

\begin{figure}
\begin{center}
\includegraphics[width=120mm]{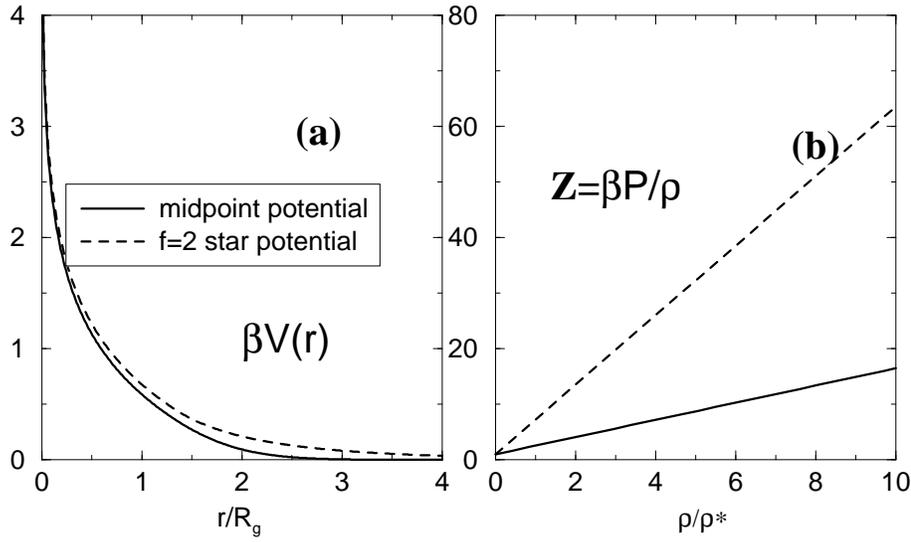}
\begin{minipage}{12cm}
\caption{\label{Fig3.4} (a) Comparison of two very similar looking potentials
that describe the midpoint--midpoint interaction. (b) Note how different
their EOS are! }
\end{minipage}
\end{center}
\end{figure}

\begin{figure}
\begin{center}
\includegraphics[width=120mm]{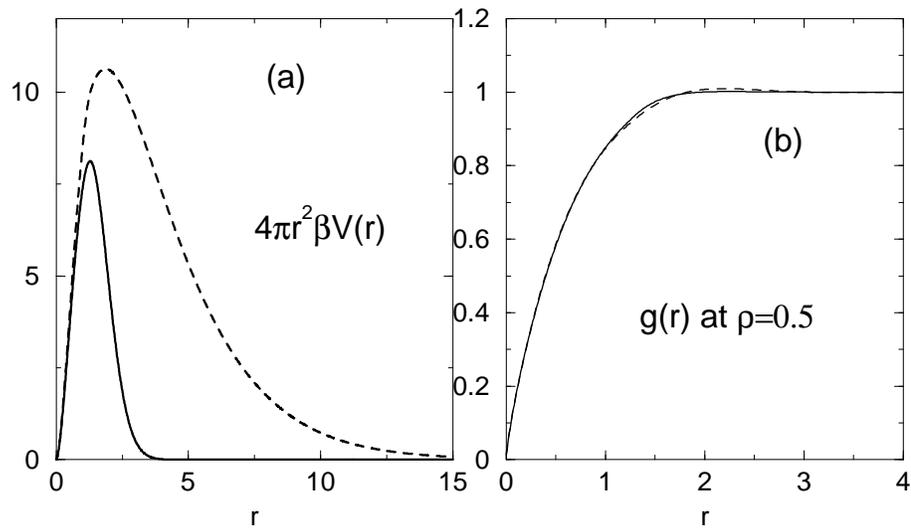}
\begin{minipage}{12cm}
\caption{\label{Fig3.5} (a) The differences between the two mid-point
potentials in the previous figure become much more apparent when they
are multiplied by $r^2$. (b) This does not translate into much difference in
the $g(r)$ shown here for  $\rho = 0.5$, but $S(k)$ (not depicted
here) does change.   }
\end{minipage}
\end{center}
\end{figure}

\begin{figure}
\begin{center}
\includegraphics[width=120mm]{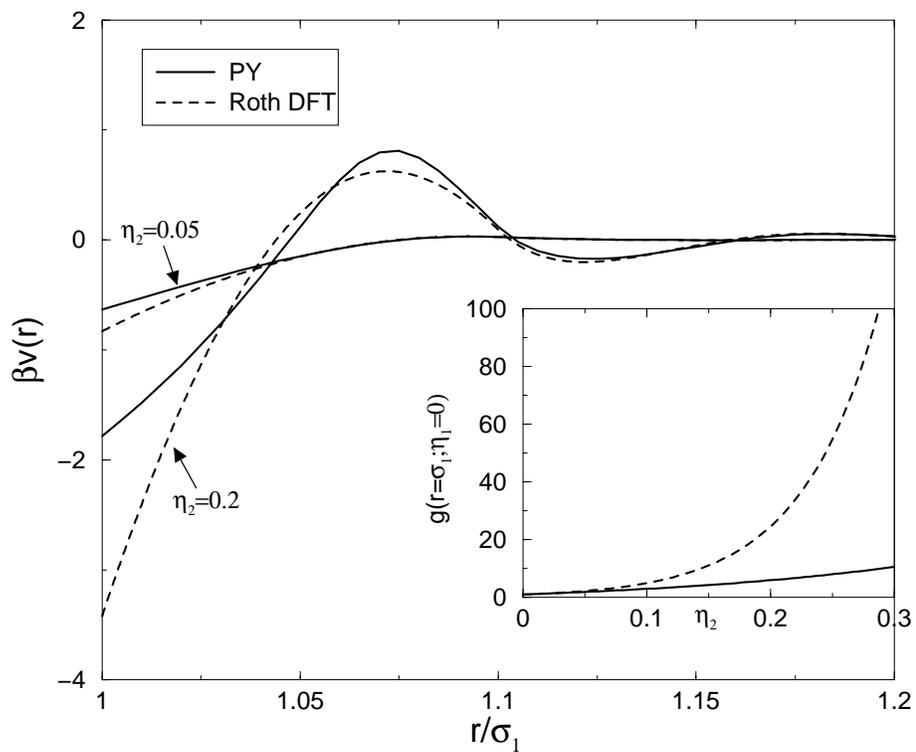}
\begin{minipage}{12cm}
\caption{\label{FigA.1} Comparison of the depletion potential
$w^{(2)}(r;\rho_2)$ derived from the PY approximation and from the
quantitatively accurate Roth DFT method discussed in the text.  PY
tends to underestimate the potential strength, leading to a much
reduced contact value of $g_{11}(r)$ in the $\rho_1 \rightarrow 0$
limit, as shown in the inset.  }
\end{minipage}
\end{center}
\end{figure}

\end{document}